\documentclass{ws-p8-50x6-00}

\def\pscz{{\it PSC}z}
\def\iras{{\it IRAS}}
\def\etal {\rm {\it et al.}}
\def\jy{1.2-Jy}
\def\markiii{{\it Mark}\,III}
\def\hm{\ifmmode\,{\it h }^{-1}\,{\rm Mpc }\else $h^{-1}\,$Mpc\,\fi}
\def\kms{\ifmmode\,{\rm km}\,{\rm s}^{-1}\else km$\,$s$^{-1}$\fi}

\begin{document}
 
\title{PSCz {\it vs.} 1.2 Jy velocity fields: a Spherical Harmonics Comparison}

\author{L. Teodoro$^1$}
\address{$^1$Instituto Superior T\'{e}cnico, Lisboa, Portugal\\
lteodoro@glencoe.ist.utl.pt}

\maketitle

\abstracts{ 
We perform a detailed comparison of the \iras\ \pscz\ and \jy\  spherical
harmonic coefficients of the predicted velocity fields in redshift
space. The monopole terms predicted from the two surveys show some
differences. Faint galaxies are responsible for this mismatch 
that disappears when extracting a \pscz\ subsample of
galaxies with fluxes larger than \jy. The analysis of \pscz\ dipole
components confirms the same inconsistencies found by Davis, Nusser and
Willick \cite{Davis:1996} between the \iras\ \jy\ gravity
field and \markiii\ peculiar velocities.
 Shot-noise, which is greatly reduced in our \pscz\ gravity field, cannot
be responsible for the observed mismatch. 
}

\section{Introduction}
Nusser \& Davis (NS)\cite{Nusser:1994} show that in linear
gravitational instability (GI) theory the peculiar velocity field in
redshift space is irrotational and thus can be expressed in terms of a
potential: ${\vec v}=-\nabla \Phi(\vec s)$. The angular
dependencies of the potential velocity field and the galaxy
overdensity field [both measured in redshift space and expanded in
spherical harmonics, $\Phi_{lm}(s)$ and $\hat{\delta}_{lm}(s)$,
respectively] are related by a modified Poisson equation:
\begin{equation}
{{1}\over{s^2}}{{d}\over{ds}}
\left( s^2 {{d\Phi_{lm}}\over{ds}} \right)
-{{1}\over{1+\beta}}
{{l(l+1)\Phi_{lm}}\over{s^2}}=
{{\beta}\over{1+\beta}}
\left( \hat{\delta}_{lm}-
{{1}\over{s}}{{d\ln{\phi}}\over{d \ln{s}}}
{{d \Phi_{lm}}\over{ds}} \right),
\label{eq:ndchapter5}
\end{equation}
where $\phi(s)$ is the selection function and $\beta \equiv \Omega ^{0.6}/b$, where $\Omega$ 
is the matter density and $b$ the bias of the galaxy distribution. To solve this differential
equation we first compute the density field on an angular grid
using cells of equal solid angle and
52 bins in redshift out to $s=18\,000$~kms$^{-1}$.
\begin{equation}
1+{\hat \delta}_j({\vec s}_n)=\frac{1}{(2\pi)^{3/2}\sigma_{1.2 n}^3}\sum_i^{N_j} \frac{1}{\phi(s_i)}\exp\left [
  -\frac{({\vec s}_n-{\vec s}_i)^2}{2\sigma_{1.2 n}^2} \right ]
\label{eq:smoothing5}
\end{equation}
where the sum is over all the galaxies within the catalogue $j$,
$N_j$.  The Gaussian smoothing width for the cell $n$ at redshift
$s_n$, $\sigma_{1.2 n}$, is given by $\sigma_{1.2 n} = \mbox{max}\{100,
[\bar{n}_{1.2}\phi_{1.2}(s_n)]^{-1/3}\}$\ km s$^{-1}~$, where $\bar{n}_{1.2}$~and
$\phi_{1.2}$~are the 1.2 Jy mean number density and selection
function, respectively.

\section{Datasets}
\label{section:datasets}

The  \iras\ \pscz\ catalogue has been recently completed
and contains 15\,500 \iras\ PSC galaxies with a 60
$\mu$m flux larger than 0.6 Jy. The average depth of this survey is $\approx$ 100
\hm.  In our analysis we will restrict
to the \pscz\ sub-sample of 11\,206 galaxies within 20\,000 \kms\ from the Local Group. 
The unobserved region is modeled by the angular mask of Saunders \etal\ \cite{Saunders:2000},
which leaves unmasked 80$\%$ of the sky. This region is 
filled-in with the cloning procedure described in Branchini 
\etal\ (B99).\cite{Branchini:1999}

The 1.2-Jy catalogue (Fisher \etal\ \cite{Fisher:1995}) contains 5\,321 \iras\ PSC
galaxies with a 60 $\mu m$ flux limit of 1.2-Jy within 20\,000 \kms\ of the Local Group. 
This catalogue has 
a slightly larger sky coverage of  $\approx$\ 87.6$\%$ and a smaller 
median distance of $\approx$ 84 $\hm$. This catalogue is
supplemented with ``synthetic'' objects in the ZoA 
and other excluded regions following the same technique.

\section{Results, Discussion and Conclusions}
\label{section:conclusions5}
\begin{figure}
\begin{center}
\scalebox{0.70}{\includegraphics[37,472][538,755]{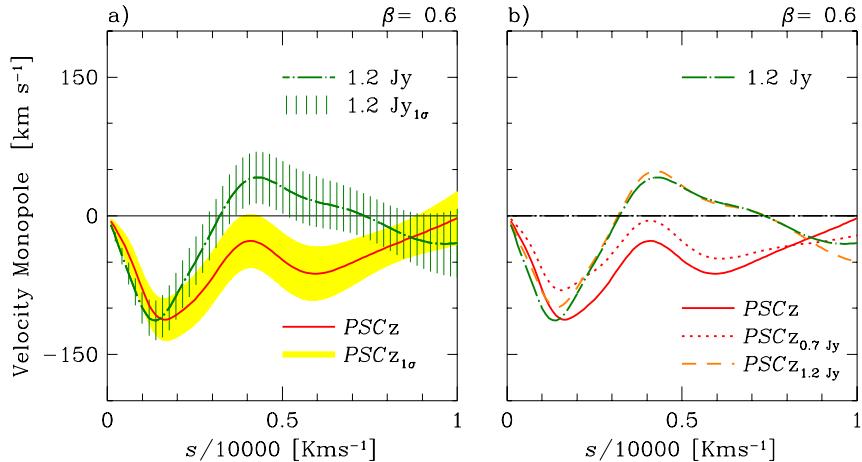}}
\caption{Monopole coefficient, $u_{00}(s)$, 
 estimated from \pscz~and 1.2-Jy catalogues for $\beta =0.6$. On the
 two top panels dashed (thick) lines represent 1.2-Jy (\pscz)
 monopole mode of the velocity field.  a) The hatched and shaded
 regions indicate the 1.2 Jy and \pscz\ 1--$\sigma$\
 uncertainties, respectively. b) Dotted, dot-dashed and dashed lines
 represent \pscz$_{0.7}$, \jy\  and
 \pscz$_{\mbox{1.2}}$, respectively. }
\label{fig:irasmonopole}
\end{center}
\end{figure}

We compare the line of sight peculiar velocities relative to the Local 
Group, $u ({\vec s})\equiv [ {\vec
v}({\vec v}) -{\vec v}_{LG} ] \cdot {\hat s}$, obtained from the \iras\ \pscz\ and 
\jy\ redshift surveys gravity fields. The comparison is performed in 
redshift-space.  To perform the decomposition in spherical
harmonics coefficients we apply $\beta=0.6$.

In Fig.~\ref{fig:irasmonopole} we display the monopole of the velocity
field, $u_{00}$. In the left panel, the
estimate of the velocity monopole in the \jy\ (dashed line) is
systematically larger than the \pscz\ one (continuous line) in the range $1\,000 < s   
 < 8\,000 $~km s$^{-1}$.

\begin{figure}
\begin{center}
\scalebox{0.55}{\includegraphics[24,268][526,690]{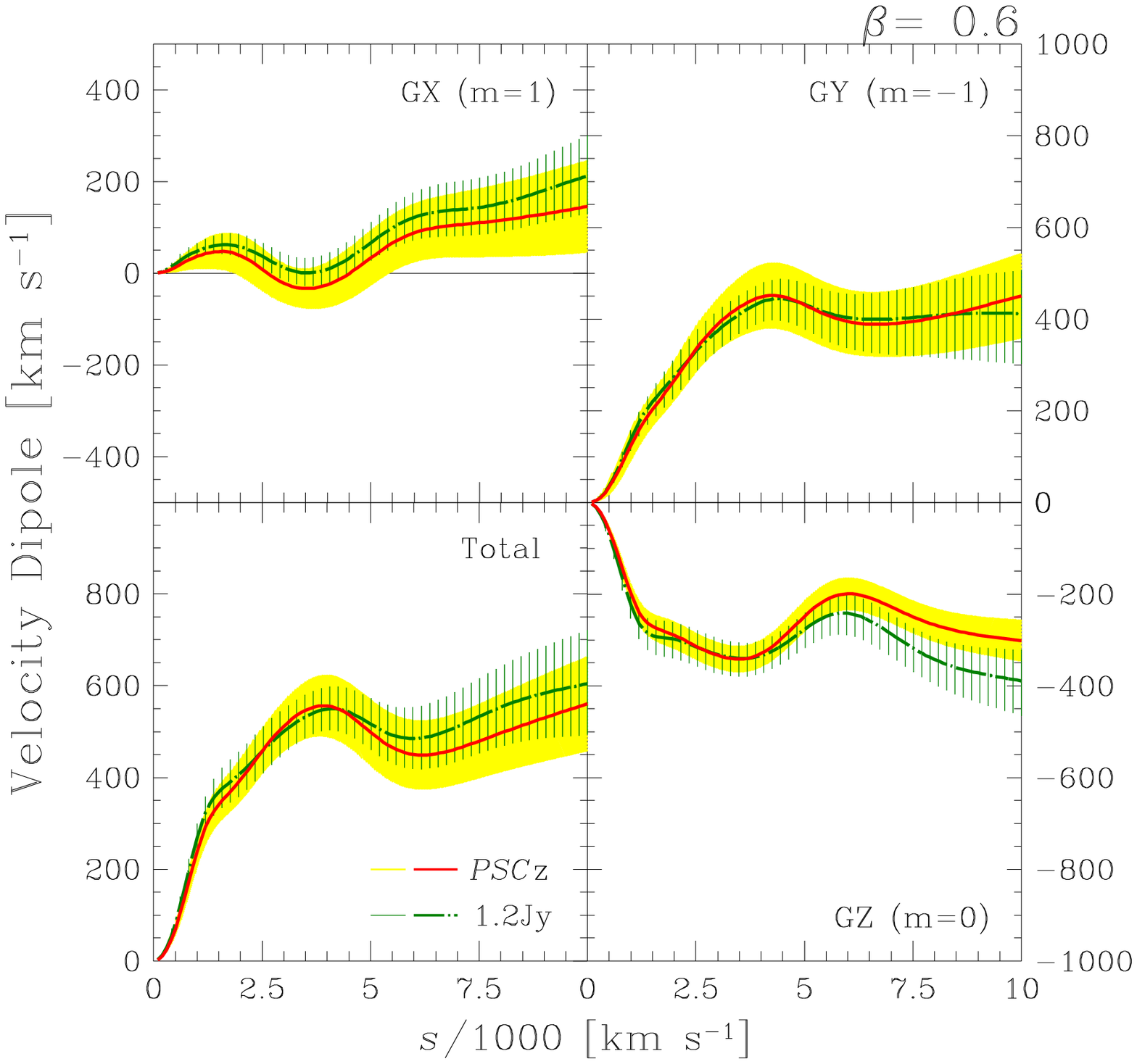}}
\caption{ The dipole coefficients, $u_{1m}(s)$\{$m=-1,0,1$\}, inferred from \pscz\ and \jy\ 
for $\beta =0.6$. The solid and dashed lines in the various panels
represent \pscz\ and \jy\ dipoles, respectively. The $GX$, $GY$,
$GZ$~panels show the three Galactic components of the dipole and the
bottom left is their quadrature sum. Hatched and shaded regions
indicate the \jy\ and \pscz\ 1--$\sigma$\ uncertainty,
respectively.}
\label{fig:irasdipoles}
\end{center}
\end{figure}

 In Fig.~\ref{fig:irasdipoles} we show the three velocity
dipole components of the two surveys along with the total amplitude (bottom
left panel). The various dipole components exhibit good agreement,
specially $u_{1,-1}$ (top right panel). Davis, Nusser and Willick (DNW)\cite{Davis:1996} 
find systematic discrepancies between the \markiii\  and \iras\ \jy--predicted 
flow fields, particularly a difference in the  $m=-1$ component of the dipole. Estimating 
the velocity field from the \iras\ \pscz\ catalogue allows us to understand the influence of 
sparse-sampling on the \jy\ inferred fields. Since both dipoles are so similar, 
we might  conclude that the discrepancies with the \markiii\ velocities 
do not disappear when the shot noise is reduced by a factor $\approx  \sqrt 3$, 
like in the \pscz\  velocity field.

Willick \etal\ applied their VELMOD machinery to compare the observed velocities 
of the \markiii\ galaxies with those predicted by the \iras\ \jy\ model 
within $ s = 3\,000$\  $\kms$. They found that the residuals are well modeled by a quadrupole
with an amplitude of $\sim 3 \%$\ that of the Hubble flow.

 In Fig. 3, the quadrupole components inferred from both \iras\ 
catalogues agree quite well, except that the $m=0,2$ magnitudes are somewhat
larger in \jy, therefore a quadrupole mismatch between \markiii\ and the 
\pscz\ velocities is also to be expected. Note that when all the 
multipoles are considered ({\it{i.e.}}\ when performing a full v-v comparison) the
\pscz\  and \jy\  gravity fields look fully consistent (B99).\cite{Branchini:1999}

In all plots, shaded and hatched regions represent 1-$\sigma$
uncertainties. These incorporate sources of errors in the velocity field: {\it i)} 
the shot-noise error due to sparse-sampling of the underlying density and velocity 
fields; {\it ii)} systematic and random errors caused by a less than perfect method of 
reconstruction. The uncertainties are evaluated summing in quadrature the 
errors due to shot-noise and filling-in procedure.  We compute the
shot-noise error by generating 100 bootstrap realizations 
of the observed distributions of \iras\ galaxies and computing the 
velocity fields from these realizations. To quantify random and systematic errors 
caused by the filling-in procedure we use a suite of 20 mock-catalogues 
that mimic the main properties of the \pscz\ and \jy\ redshift surveys.
A complete description of the  prescription 
followed to quantify the velocity uncertainties can be found in 
Teodoro, Branchini and Frenk \cite{Teodoro:2000}.

Where does the discrepancy
between the monopoles of the \pscz\ and \jy\ surveys come from? As
shown in the plots, the difference is larger than that expected from the
shot--noise (included in the error budget). 
Tadros {\it et al.}\cite{Tadros:1999} have
suggested that the \pscz\ catalogue may be incomplete for fluxes $\le
0.7$~Jy. If true, then we would expect that the velocity monopole for
the \pscz\ with a flux cut at 0.7 Jy (\pscz$_{0.7}$) would be in good
agreement with the 1.2-Jy survey. The dotted line
($u_{00,\mbox{\pscz}_{0.7}}$) in the top panel of
Fig. \ref{fig:irasmonopole} shows shows that objects fainter than 0.7 Jy 
are only partially responsible for the monopole  mismatch that therefore 
cannot be ascribed 
to incompleteness at low fluxes however that this is not case. 
It is only
in cutting the \pscz\ catalogue at a flux level of 1.2 Jy that, as
expected, the discrepancy disappears.
 This is clearly seen in the right panel
of Fig. \ref{fig:irasmonopole} in which the dashed line
indicates  \pscz$_{1.2}$ velocity monopole.

We conclude that a better sampled \iras\ catalogue cannot resolve alone 
the mismatches between the \markiii\ peculiar velocity field and 
the \iras\ predicted gravity field.

\begin{figure}
\begin{center}
\scalebox{0.55}{\includegraphics[24,27][526,766]{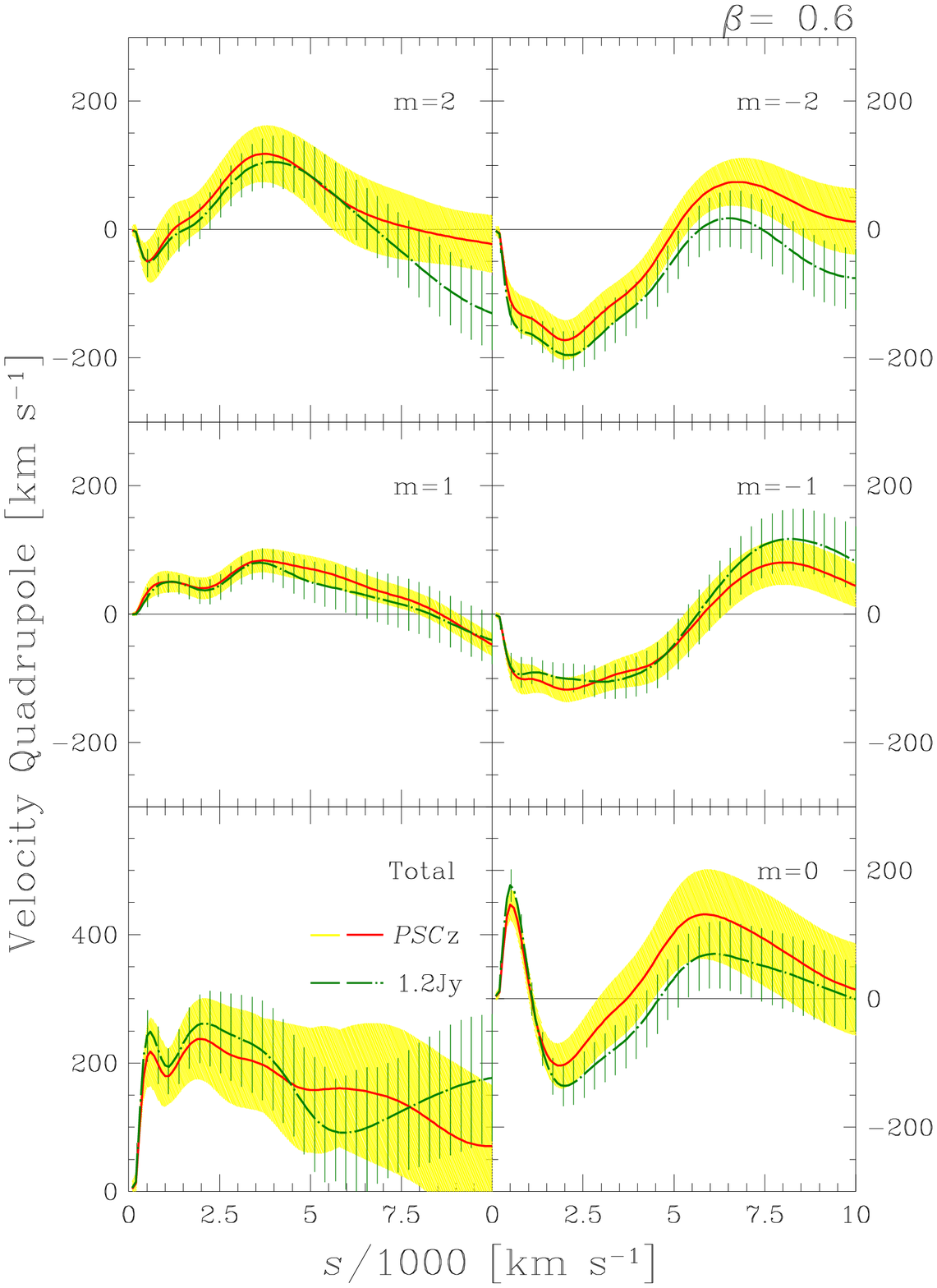}}
\caption{ The quadrupole coefficients, $u_{1m}(s)$\{$m=-2,-1,0,1,2$\}, 
inferred from \pscz\ and 1.2-Jy
for $\beta =0.6$. The solid and dashed lines in the various panels
represent \pscz\ and 1.2-Jy quadrupoles, respectively. The $m = 0,\pm 1, \pm 2 $\ panels 
show the five Galactic components of the quadrupole and the
bottom left is their quadrature sum. Hatched and shaded regions
indicate the 1.2-Jy and \pscz~1--$\sigma$\ uncertainty,
respectively.}
\label{fig:irasquadrupole}
\end{center}
\end{figure}

\section*{ Acknowledgments}
LT has been supported by the grants PRAXIS XXI/BPD/16354/98 and 
PRAXIS/C/FIS/13196/1998. Thanks do Enzo Branchini and Carlos Frenk for advice on this 
work as well as comments on the text.

\end{document}